\documentclass[aps, superscriptaddress]{revtex4-1}

\usepackage[dvipdfmx]{graphicx}
\usepackage{amsmath}
\usepackage{amssymb}
\usepackage{bm}
\usepackage{color}

\begin{document}
	\title{Noise-tuned bursting in a Hedgehog burster}
	\author{Jinjie Zhu}
	\email{zhu.j.ag@m.titech.ac.jp}
	\affiliation{School of Mechanical Engineering, Nanjing University of Science and Technology, Nanjing 210094, China}
	\affiliation{Department of Systems and Control Engineering, Tokyo Institute of Technology, Tokyo 152-8552, Japan}
	\author{Hiroya Nakao}
	\affiliation{Department of Systems and Control Engineering, Tokyo Institute of Technology, Tokyo 152-8552, Japan}
	\date{\today}
	
\begin{abstract}
Noise can shape the firing behaviors of neurons. Here, we show that noise acting on the fast variable of the Hedgehog burster can tune the spike counts of bursts via the self-induced stochastic resonance (SISR) phenomenon. Using the distance matching condition, the critical transition positions on the slow manifolds can be predicted and the stochastic periodic orbits for various noise strengths are obtained. The critical transition positions on the slow manifold with non-monotonic potential differences exhibit a staircase-like dependence on the noise strength, which is also revealed by the stepwise change in the period of the stochastic periodic orbit. The noise-tuned bursting is more coherent within each step while displaying mixed-mode oscillations near the boundaries between the steps. When noise is large enough, noise-induced trapping of the slow variable can be observed, where the number of coexisting traps increases with the noise strength. It is argued that the robustness of SISR underlies the generality of the results discovered in this paper.
	%%% Leave the Abstract empty if your article does not require one, please see the Summary Table for full details.
\end{abstract}

	\maketitle

\section{Introduction}

How can we change the spike counts of a bursting neuron? From the viewpoint of dynamical systems, a burst consists of a series of spikes followed by a period of quiescence, and the spiking activity is caused by the fast currents while modulated by the slow currents through a variety of bifurcations \cite{Izhikevich2007}. For a single unit, varying the bifurcation parameter or the timescale separation parameter may lead to changes in spike counts of a burst \cite{Wang1993,Barrio2011,Zhu2017}. For coupled bursting neurons, the effects of time delay and coupling can also induce various bursting patterns with different numbers of spikes in each burst \cite{Dhamala2004,Jia2018,Zhu2017} or even termination of bursting such as amplitude death \cite{Thottil2017}.

Noise is inevitable and may play functional roles in neuronal systems \cite{Lindner2004,McDonnell2011,Faisal2008,Schmerl2013,Bauermann2019}. For excitable neurons close to bifurcation, noise of moderate strength can give rise to coherent oscillations. This noise-induced oscillatory phenomenon was named coherence resonance (CR) \cite{Gang1993,Longtin1997,Pikovsky1997}. For noise acting on the fast variable, Muratov {\it et al.} \cite{Muratov2005,LeeDeVille2005} found a stochastic resonance-like coherent behavior and referred to it as self-induced stochastic resonance (SISR). Both CR and SISR can be significantly relevant to the dynamical behaviors in neuronal systems \cite{Yamakou2019,Touboul2020,Yamakou2020FiCN,Baspinar2021,Zhu2022PRR}. In particular, SISR is proven to be more robust than CR since SISR does not require the system to be close to bifurcation \cite{LeeDeVille2005}. Further, SISR can occur in a wider range of noise strengths and different noise strengths can induce different coherent stochastic orbits \cite{LeeDeVille2005,Zhu2021PRR}. This property is highly relevant to our question, that is, noise may also tune the spike counts in a bursting neuron.

The mathematical mechanism of the SISR oscillator was first analyzed by Freidlin regarding the motion of a light particle in the force field perturbed by small noise \cite{Freidlin2001}. Later, Muratov {\it et al.} showed its ubiquity in randomly perturbed excitable systems and termed it as SISR \cite{Muratov2005}. The large timescale separation is important for the realization of SISR in that larger timescale separation enables larger intervals of noise strengths for coherence to occur~\cite{Muratov2005,Yamakou2018}. In the limit of vanishing noise and large timescale separation of fast-slow systems, the stochastic dynamics exhibit almost deterministic periodic oscillations, which can be predicted by the timescale matching condition \cite{Muratov2005,LeeDeVille2005,Yamakou2018,Yu2021}. However, this condition does not consider the differences in timescales on different branches, which can be important to the transition positions, e.g., in the FitzHugh-Nagumo (FHN) neuron model~\cite{Zhu2021PRR,Zhu2022PRR}. Recently, we proposed a distance matching condition, which solves this problem by defining the so-called mean first passage velocity \cite{Zhu2021PRR}. In this paper, we use this condition to analyze the bursting neuron whose potential difference between the slow manifold and the barrier is non-monotonic. Our analysis reveals the possibility to control the spike counts in the bursting neurons solely by the noise strength.

\section{Materials and methods}

\subsection{Hedgehog burster}
\begin{figure}
	\centering
	\includegraphics[width=1\textwidth]{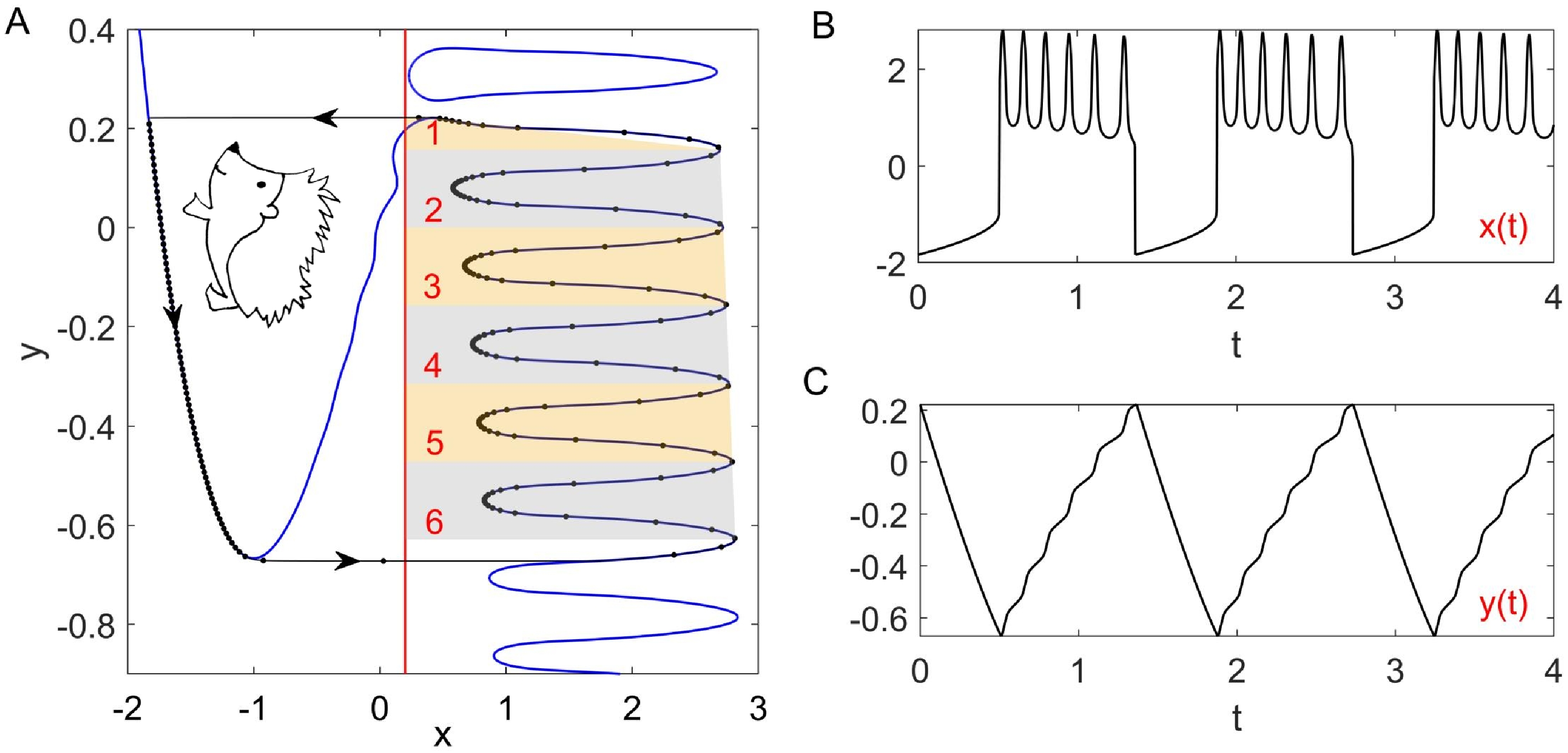}
	\caption{Phase portrait and time series of the Hedgehog burster (\ref{eq:1}). (A) Phase portrait. The blue curves and the red line represent the $x$ and $y$ nullclines, respectively. The black curve with arrows illustrates the stable limit cycle, where the dots indicate the states on the limit cycle with equal time intervals. (B,C) Time series of the membrane potential $x(t)$ and the recovery variable $y(t)$.}
	\label{fig:1}       % Give a unique label
\end{figure}

We consider a modified version of the FHN neuron model, which can exhibit bursting behaviors in two dimensions. This model is named Hedgehog burster by Izhikevich \cite{Izhikevich2000} due to the hedgehog-like limit-cycle attractor. The governing equations are as follows:
\begin{equation}
	\begin{split}
		\varepsilon \frac{\mathrm{d}x}{\mathrm{d}t} &= f(x,y),\\
		\frac{\mathrm{d}y}{\mathrm{d}t} &= g(x,y),
	\end{split}
	\label{eq:1}
\end{equation}
where $f(x,y) = x-\frac{x^3}{3}-y+4\,L(x)\cos(40y)$, $g(x,y) = x+a$, and $x$, $y$ represent the fast membrane potential and the slow recovery variable, respectively. The timescale separation is characterized by the small parameter $\varepsilon=0.0001$. The bifurcation parameter $a=-0.2$ is fixed so that the fixed point is unstable and there is a stable limit cycle as shown in Figure~\ref{fig:1}A. The Hedgehog burster described by Eq.~(\ref{eq:1}) is different from the original FHN model because of the cosine term $\cos(40y)$ with the logistic function $L(x)=\frac{1}{1+\exp[5(1-x)]}$. The logistic function $L(x)$ is chosen so that the left branch of the $x$ nullcline remains nearly unchanged while the right branch fluctuates, which is key to the realization of the bursting behavior. The large timescale separation forces the Hedgehog limit cycle to stick to the slow manifolds (the left and right branches of the $x$ nullcline). As such, the wavy right branch of the $x$ nullcline determines the spike counts in each burst (six spikes in Figure~\ref{fig:1}B), which can be also observed in the time series of $y(t)$ as in Figure~\ref{fig:1}C. It can be observed that there is a closed loop of the $x$ nullcline on top of the right branch (In fact, there are more such loops above). Initial states on it can also stick to the right half of the loop and may give a spike. However, the dynamics related to this loop are mainly transient, which are thus out of the scope of this paper.

\subsection{Stochastic periodic orbits by SISR}
Noise can not only modify the system's deterministic behavior, but can also induce new phenomena that are not observed in deterministic systems. Muratov {\it et al.} \cite{Muratov2005,LeeDeVille2005} showed that noise acting on the fast subsystem can give rise to a stochastic resonance-type phenomenon, which was named SISR. In SISR, the system's state follows the slow manifold while making an almost deterministic transition at a critical position so that the stochastic oscillation can be very coherent. SISR resembles the stochastic resonance (SR) phenomenon in that both SR and SISR exhibit barrier-crossing behaviors. Besides, the potential difference is modulated by the external periodic signal in SR while by the slow variable in SISR.

The noise-induced coherent orbits caused by SISR are completely different from the deterministic limit cycle, as we will see later for the Hedgehog burster. Here, we show for the Hedgehog burster with non-monotonic potential differences, the SISR can still be predicted by using our previously proposed distance matching condition~\cite{Zhu2021PRR}.

The governing equation for the noise-perturbed Hedgehog burster is as follows:
\begin{equation}
	\begin{split}
		\varepsilon \frac{\mathrm{d}x}{\mathrm{d}t} &= f(x,y)+ \sqrt{\varepsilon} \xi(t),\\
		\frac{\mathrm{d}y}{\mathrm{d}t} &= g(x,y),
	\end{split}
	\label{eq:2}
\end{equation}
where $f(x,y)$ and $g(x,y)$ are the same as in Eq.~(\ref{eq:1}), $\xi(t)$ is the Gaussian white noise satisfying $\langle\xi(t)\rangle=0$, and $\langle\xi(t)\xi(t')\rangle=\sigma \delta(t-t')$, where $\sigma$ represents the noise strength. Figures \ref{fig:5} and \ref{fig:6} illustrate typical noise-tuned bursting dynamics of the system~(\ref{eq:2}) obtained by the Monte Carlo simulations with several noise strengths. Depending on the noise strength, the system exhibits different stochastic periodic orbits. To theoretically predict the stochastic periodic orbit of the SISR phenomenon, we follow the similar procedure as in Ref.\cite{Zhu2021PRR}. 

\section{Results}

\subsection{Noise-tuned bursting in hedgehog burster}
In the following, we will apply the distance matching condition~\cite{Zhu2021PRR} to predict the transition positions on the stable branches of the $x$ nullcline. The distance matching condition compares the noise-induced displacement of the state away from the stable branch and the distance from the stable branch to the unstable one. When these two distances are equal to each other, transition happens with a large probability.

\begin{figure}
	\centering
	\includegraphics[width=1\textwidth]{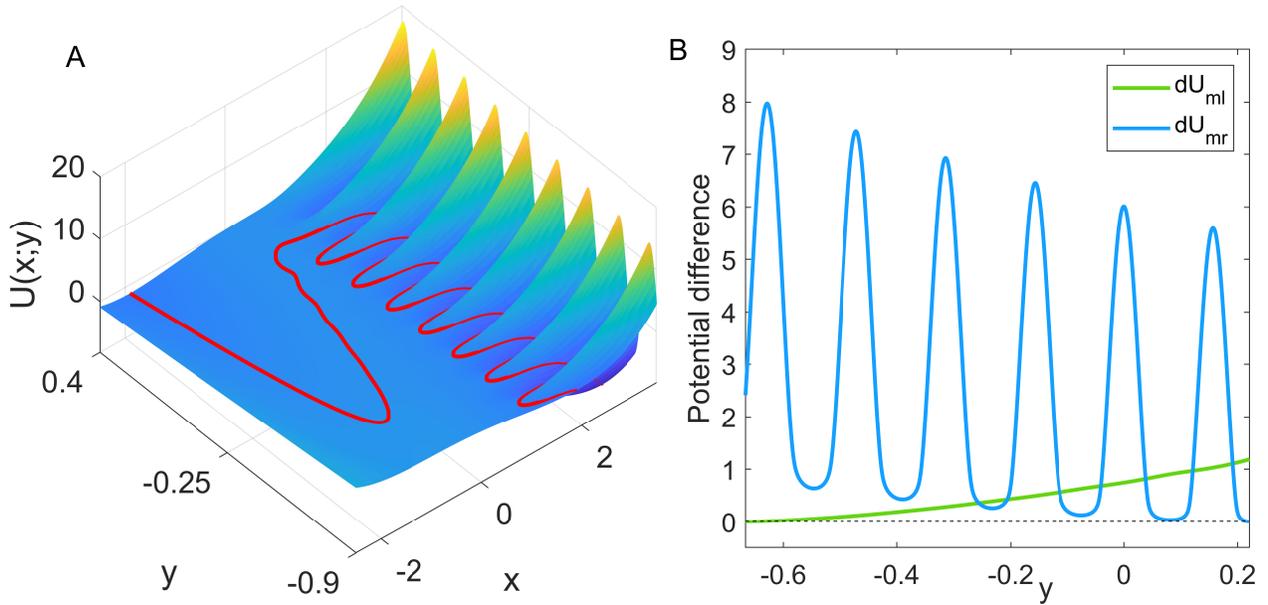}
	\caption{(A) Potential function $U(x;y)$, where $y$ is regarded as a fixed parameter. The red curve is the $x$ nullcline, which follows the local extrema of the potential for each fixed $y$. (B) Potential differences between the middle and left branches, $dU_{ml}=U_m-U_l$ and between the middle and right branches, $dU_{mr}=U_m-U_r$, where $U_l$, $U_m$, and $U_r$ denote the potential values on the left, middle, and right branches for given $y$, respectively.}
	\label{fig:2}       % Give a unique label
\end{figure}

When the state is on the left branch of the $x$ nullcline, for each fixed $y$, there is a corresponding mean first passage time (MFPT) $T_{\rm e}$ obtained from the Kramers rate \cite{Kramers1940,Gardiner1985}:
\begin{equation}
	T_{\rm e}(y)=\frac{2 \pi}{\sqrt{\lvert U''_{m}(x;y) \rvert U''_{l}(x;y)}}{\rm exp}\left(\frac{2 dU_{ml}}{\sigma}\right),
	\label{eq:3}
\end{equation}
where the potential function $U(x;y)=-\int f(x,y) \mathrm{d}x+C$ ($C$ is a constant and $y$ is regarded as a fixed parameter). The double prime over the potential function represents the second-order derivative with respect to $x$. Denoting the potential function on the left, middle, and right branches as $U_l$, $U_m$, and $U_r$, respectively, the potential difference between the middle and left branches, $dU_{ml}=U_m-U_l$, and the middle and right branches, $dU_{mr}=U_m-U_r$, can be easily calculated. The landscape of the potential function (the constant $C$ can be safely set to zero as it is eliminated in Eq.(\ref{eq:3})) and the potential differences are shown in Figure~\ref{fig:2}. The potential difference $dU_{ml}$ monotonically decreases for decreasing $y$. However, the potential difference $dU_{mr}$ largely fluctuates while decreases on average for increasing $y$. Thus, the potential differences have more than one intersection points, which is significantly different from the FHN neuron model with monotonic potential differences~\cite{Zhu2021PRR}. The intersection points in Figure~\ref{fig:2}B give the values of $y$ at which the two potential differences are equal to each other. At these points, the boundary crossing of $x$ from the left branch to the middle branch and that from the right branch to the middle branch are equally difficult. 

From the MFPT, the mean first passage velocity (MFPV) function can thus be defined as \cite{Zhu2021PRR}:
\begin{equation}
	V_{\rm e}(y)=\frac{S(y)}{T_{\rm e}(y)},
	\label{eq:4}
\end{equation}
where $S(y)=x_m(y)-x_l(y)$ is the distance between the middle and the left branches of the $x$ nullcline for fixed $y$. The distance matching condition for the transition from the left branch to the middle branch is as follows: for the system state starting from $y_0(t_0)$ and reaching the critical transition position $y_*(t_*)$, the total displacement in the $x$ direction from the left branch can be obtained by integrating the MFPV over time, which should be equal to the distance $S(y)$ between the branches. By using Eqs. (\ref{eq:3}) and (\ref{eq:4}), we can express the distance matching condition in a self-consistent manner \cite{Zhu2021PRR} as
\begin{equation}
	\int^{y_{*}}_{y_{0}} \frac{S(y)\sqrt{\lvert U''_{m}(x;y) \rvert U''_{l}(x;y)}}{2 \pi \left(x_l(y)+a\right)\varepsilon \, {\rm exp}\left(\frac{2 dU_{ml}}{\sigma}\right)}\,\mathrm{d}y=S(y_{*}),
	\label{eq:5}
\end{equation}
where we have changed the variable of integration from $t$ to $y$ by using $\mathrm{d}y = \varepsilon( x + a ) \mathrm{d}t$. By solving Eq.~(\ref{eq:5}), we can obtain the critical transition position $y_{*}$. 

\begin{figure}
	\centering
	\includegraphics[width=1\textwidth]{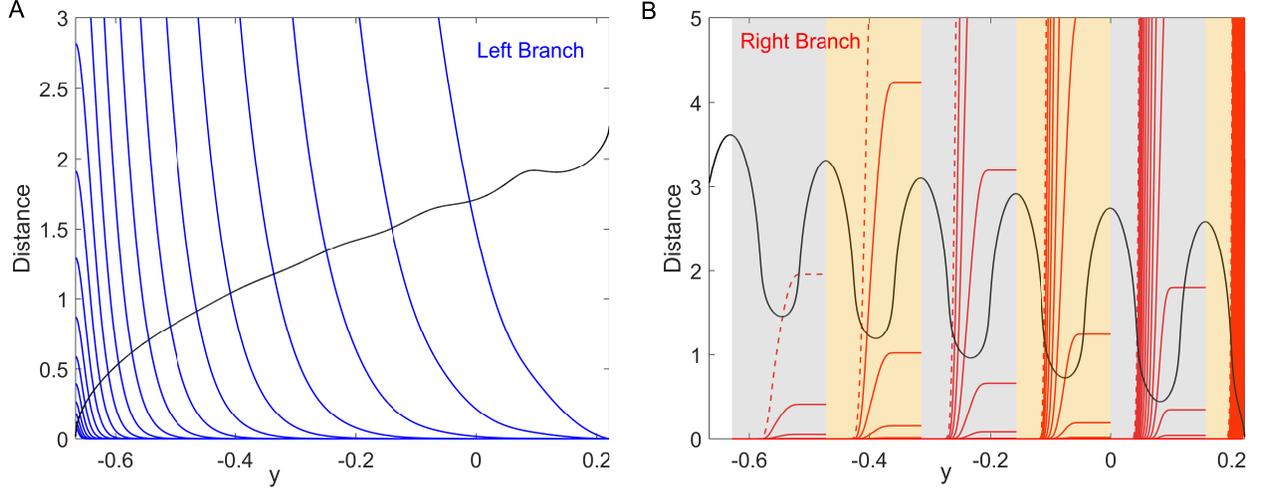}
	\caption{Identification of the critical transition positions by the distance matching condition, Eq.~(\ref{eq:5}). (A) Left branch. The blue curves and the black curve represent the lhs and rhs of Eq.~(\ref{eq:5}), respectively. (B) Right branch. The red curves and the black curve represent the lhs and rhs of Eq.~(\ref{eq:5}), respectively. Six patches in gray and yellow correspond to those in Figure~\ref{fig:1}A. Noise strengths for the integration curves of the lhs of Eq.~(\ref{eq:5}) are from $10^{-3}$ to $10^{-0.5}$ with equal logarithmic intervals ((A): from left to right; (B): from right to left). The red dashed curves in (B) display the result for $\sigma=10^{-0.5}$.}
	\label{fig:3}
\end{figure}

\begin{figure}
	\centering
	\includegraphics[width=1\textwidth]{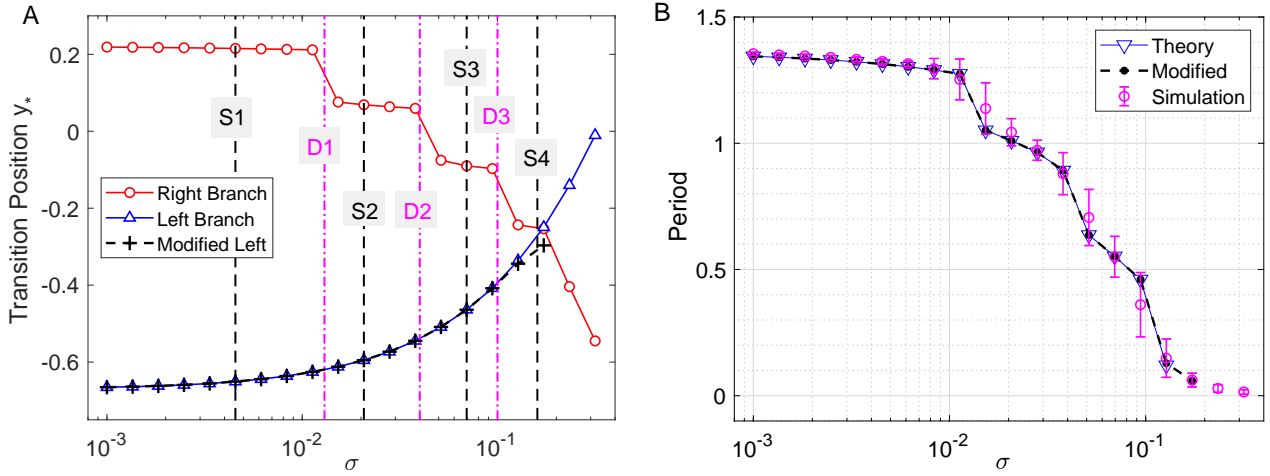}
	\caption{Critical transition positions and bursting periods. (A) Critical transition positions on the left branch (blue triangles) and right branch (red circles). The black crosses represent the transition positions on the left branch obtained by modifying the lower limit of integration as explained in the text. (B) Bursting periods. Theoretical prediction, Eq.~(\ref{eq:6}) vs. the Monte Carlo simulations (averaged for $t=200$ with error bars showing the standard deviation). Noise strengths are the same as in Figure~\ref{fig:3}.}
	\label{fig:4}
\end{figure}

\begin{figure}
	\centering
	\includegraphics[width=1\textwidth]{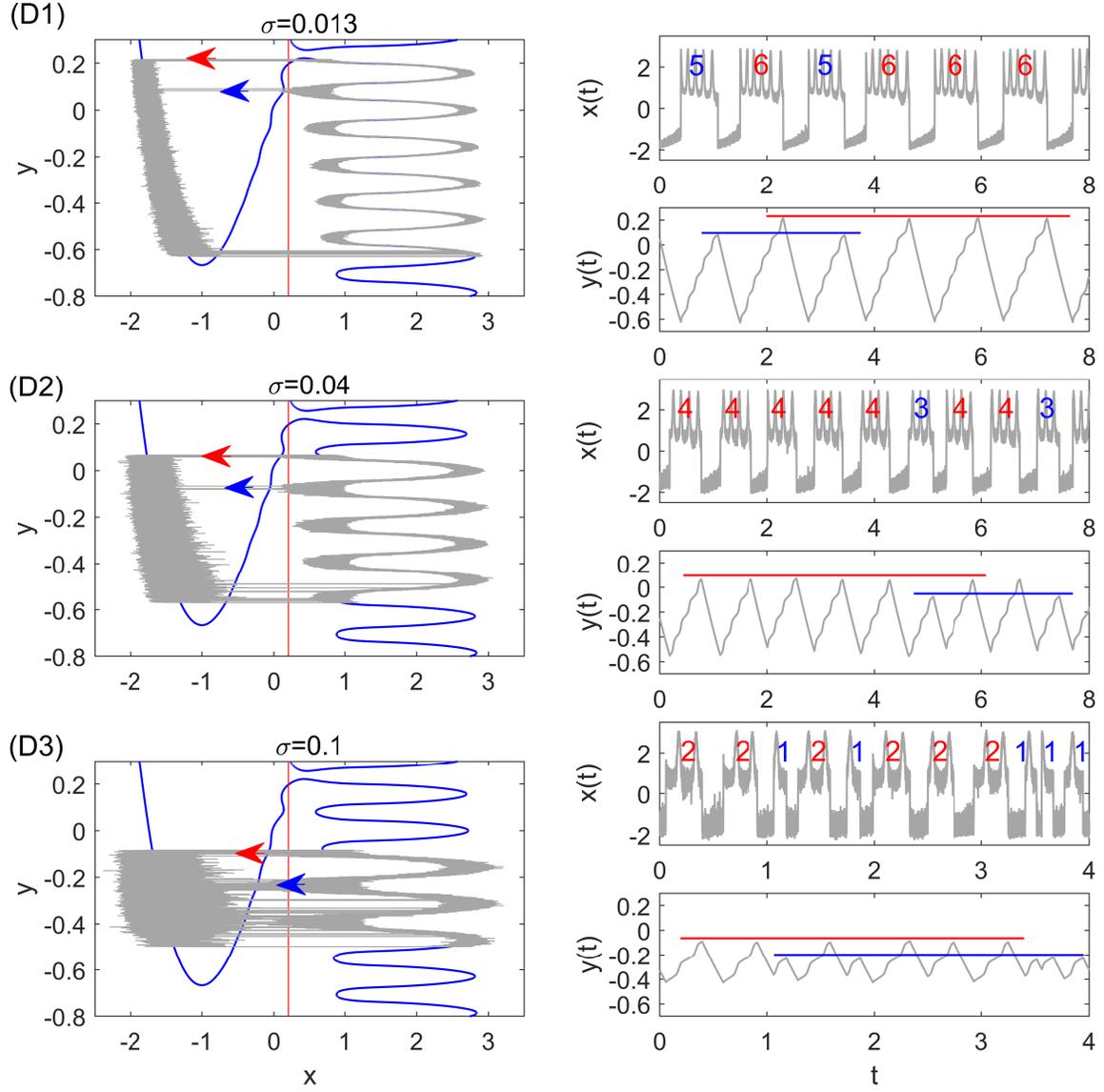}
	\caption{Stochastic periodic orbits and time series with the double-escape transition for different noise strengths. The thin gray curves are the results of Monte Carlo simulations. From top to bottom, the parameters are: $\sigma=0.013$(D1), $\sigma=0.04$(D2), and $\sigma=0.1$(D3). They correspond to the pink dot-dashed lines D1, D2, and D3 in Figure~\ref{fig:4}A. The red and blue arrows in the left panel display the two transition positions, which lead to different spike counts in the burst. The red and blue lines in the right panel illustrate mixed-mode oscillations of the slow variable $y(t)$.}
	\label{fig:6}       % Give a unique label
\end{figure}

Figure~\ref{fig:3}A plots the integral on the left-hand side (lhs) and the distance on the right-hand side (rhs) of Eq.~(\ref{eq:5}) for the left branch as functions of $y$ for different noise strengths. The intersection points between the lhs (blue) and rhs (black) correspond to the critical transition position $y_{*}$. The starting position of the system state is chosen as $y_0=0.221$, which is the maximum value of $y$ for the deterministic limit cycle on the left branch shown in Figure~\ref{fig:1}A. We will see later that the starting position is not important as long as it is far enough from $y_{*}$. The critical transition position on the left branch is plotted against the noise strength in Figure~\ref{fig:4}A (blue triangles). As expected, the transition position gradually increases with the noise strength.

The critical transition position on the right branch can be similarly calculated. However, due to the non-monotonic potential difference as shown in Figure~\ref{fig:2}, the transition process along the right branch should be treated differently. To this end, we divide the right branch into six regions as shown in Figure~\ref{fig:1}A and apply the distance matching condition separately. In each region, we start the system state from the bottom-rightmost position on the branch (where the potential is locally minimum) and seek the self-consistent solution to Eq.~(\ref{eq:5}), where the lower limit $y_0$ of integration is taken as the $y$ coordinate of the starting position. This procedure is started over again in every region. The results on the right branch are illustrated in Figure~\ref{fig:3}B for different noise strengths, where the first intersection of the lhs (red) and rhs (black) gives the critical transition position. It should be noted that the intersection takes place only on the left half of each region before the minimum of the distance curve (rhs, black). This means that if the transition cannot occur before the position with the minimum potential difference, then the transition after that position is extremely difficult in each region.

Figure \ref{fig:4}A also displays the obtained critical transition positions for various noise strengths on the right branch (red circles), which shows a significantly different tendency from those on the left branch (blue triangles). The transition positions exhibit several stages and decrease stepwisely with the noise strength, corresponding to the different regions on the right branch in Figure~\ref{fig:1}A. In Figure~\ref{fig:4}A, the transition positions on the left and right branches intersect at $(\sigma,y_*)\approx (0.173,-0.253)$, which corresponds to the case that the transitions occur at the same position ($y_{*l}=y_{*r}\approx -0.253$). Above this noise strength, i.e., $\sigma>0.173$, the predicted transition positions cannot constitute a complete orbit and are meaningless.

In numerical simulations, further increasing the noise strength ($\sigma>0.173$) will make the actual transition positions on the left and right branches asymptotically approach each other (but never cross). The crossing at $(\sigma,y_*)\approx (0.173,-0.253)$ of the predicted transition positions in Figure~\ref{fig:4}A is due to the fixed initial position $y_0$. A more appropriate value for the starting position $y_0$ is to choose the critical transition position on the other branch. We modify the starting positions on the left branch to the transition positions on the right branch and obtain the corrected $y_*$ as in Figure~\ref{fig:4}A. The modified results almost coincide with the original ones except when the two transition positions on the left and right branches get too close. This is because only a small interval of $y$ contributes to the transition process, as can be seen from the steep curves in Figure~\ref{fig:3}. Therefore, $y_0$ can be chosen arbitrarily above but not close to the transition position \cite{Zhu2021PRR}. 

\begin{figure}
	\centering
	\includegraphics[width=1\textwidth]{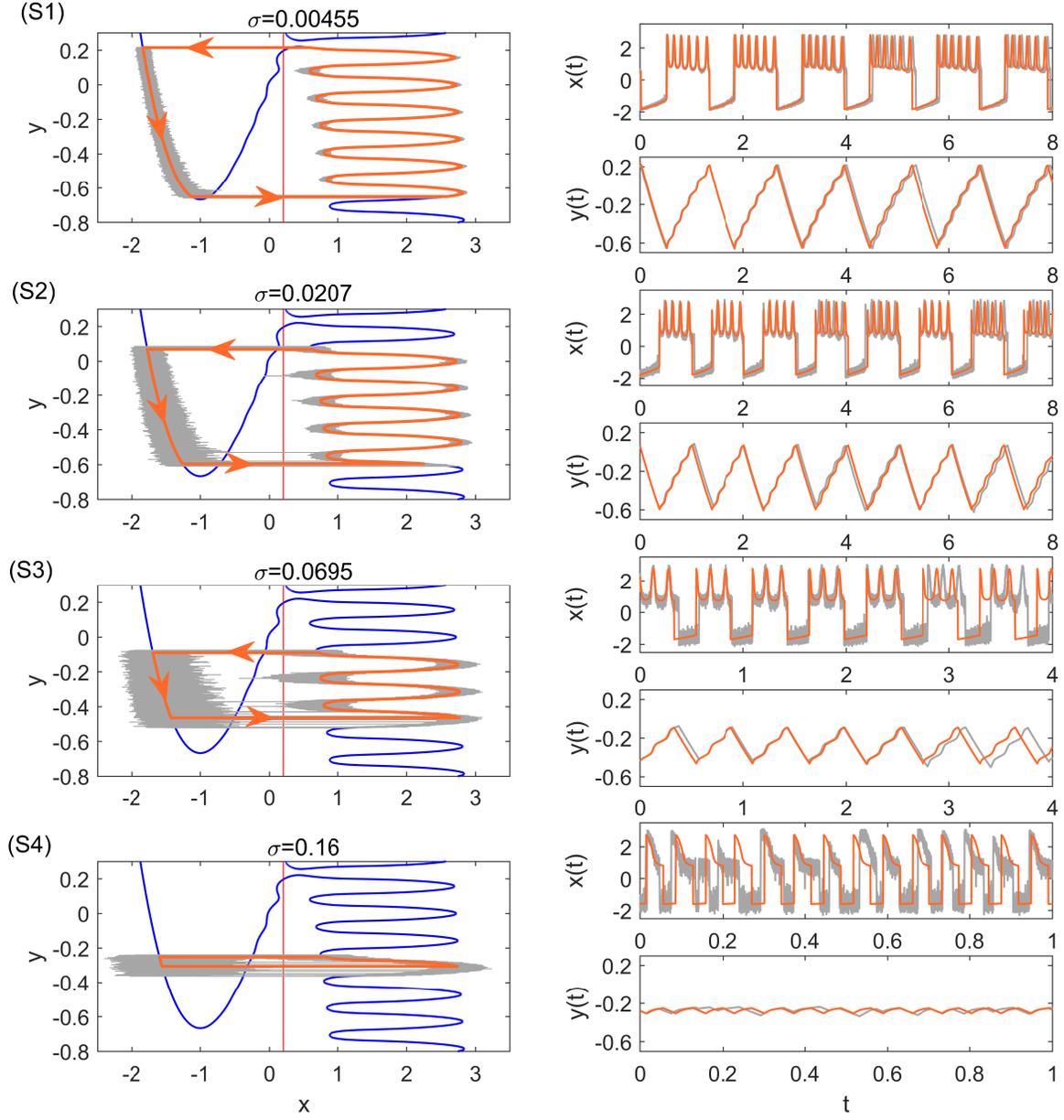}
	\caption{Stochastic periodic orbits and time series with the single-escape transition for different noise strengths. The bold orange and the thin grey curves are the theoretical periodic orbits and the results of Monte Carlo simulations, respectively. From top to bottom, the parameters are: $\sigma=0.00455$(S1), $\sigma=0.0207$(S2), $\sigma=0.0695$(S3), and $\sigma=0.16$(S4). They correspond to the black dashed lines S1, S2, S3, and S4 in Figure~\ref{fig:4}A.}
	\label{fig:5}       % Give a unique label
\end{figure}

After computing the transition positions on each branch, the complete stochastic periodic orbit can be accordingly obtained by gluing the slow motions along the slow manifolds at these transition positions. Figure~\ref{fig:5} plots the predicted stochastic periodic orbits for several representative noise strengths that correspond to the black dashed lines labeled as S1, S2, S3, and S4 in Figure~\ref{fig:4}A. The theoretical predictions are in good agreement with the results of Monte Carlo simulations. Even for large enough noise (S4), the range of $y$ can be well predicted despite the relatively poor prediction on $x$. Increasing the noise strength will decrease the size of the stochastic periodic orbit. Therefore, it can be seen that noise can tune the spike counts in every burst at different stages in Figure~\ref{fig:4}A. From S1 to S4, there are 6,5,3, and 1 spikes in each burst.

Considering the large timescale separation, the period of the bursting orbit can be approximated by the motion along the slow manifolds on the left and right branches, which is given as
\begin{equation}
	T_{\rm bursting}=\int_{y_{cr}}^{y_{cl}} \frac{\mathrm{d}y}{\varepsilon \left(x_l(y)+a\right)}+\int_{y_{cl}}^{y_{cr}} \frac{\mathrm{d}y}{\varepsilon \left(x_r(y)+a\right)},
	\label{eq:6}
\end{equation}
where $y_{cl}$ and $y_{cr}$ denote the critical transition positions on the left and right branches, respectively. The predicted periods are also in good agreement with the simulation results in different transition stages as shown in Figure~\ref{fig:4}B. The small standard deviation within each stage implies highly coherent oscillations, which can be observed in Figure~\ref{fig:5}. We can see that for each noise strength, the transition occurs near the leftmost tip of the right branch of $x$ nullcline in each region. There is only one transition position on the right branch, which we call single-escape transition. On the other hand, the period near the boundary between two different stages has a larger standard deviation. The overestimation and underestimation of the bursting period near the boundaries imply the emergence of the mixed-mode oscillations. Three typical noise-induced mixed-mode oscillations are illustrated in Figure~\ref{fig:6}. In contrast to Figure~\ref{fig:5}, the system state can transit at two tips of the right branch, which we call double-escape transition. Indeed, for the noise strength near the boundary between the two stages of coherent oscillations, the critical transition position begins to switch from one region to the other (i.e., the first intersection point switches from one valley to the other of the black distance curve in Figure~\ref{fig:3}B). Therefore, for these noise strengths, the transition can occur at different tips on the right branch for different realizations.

\subsection{Large noise-induced slow variable traps}
As discussed in the previous section, for large noise, the critical transition positions on the left and right branches approach each other asymptotically. For the FHN model, we have shown in Ref.~\cite{Zhu2021PRR} that, as the noise strength increases, the size of the stochastic periodic orbit will shrink and the period will approach zero. This is also true in the present case as we can see from Figure~\ref{fig:4}B or Figure~\ref{fig:5}. Furthermore, for sufficiently large noise, there appears a stable region where the slow variable $y$ is almost clamped at a fixed value in the steady state after the transient in the FHN system \cite{Zhu2021PRR} (see also \cite{Touboul2020} for large noise-induced asynchrony in interacting neuronal ensembles). 

\begin{figure}
	\centering
	\includegraphics[width=1\textwidth]{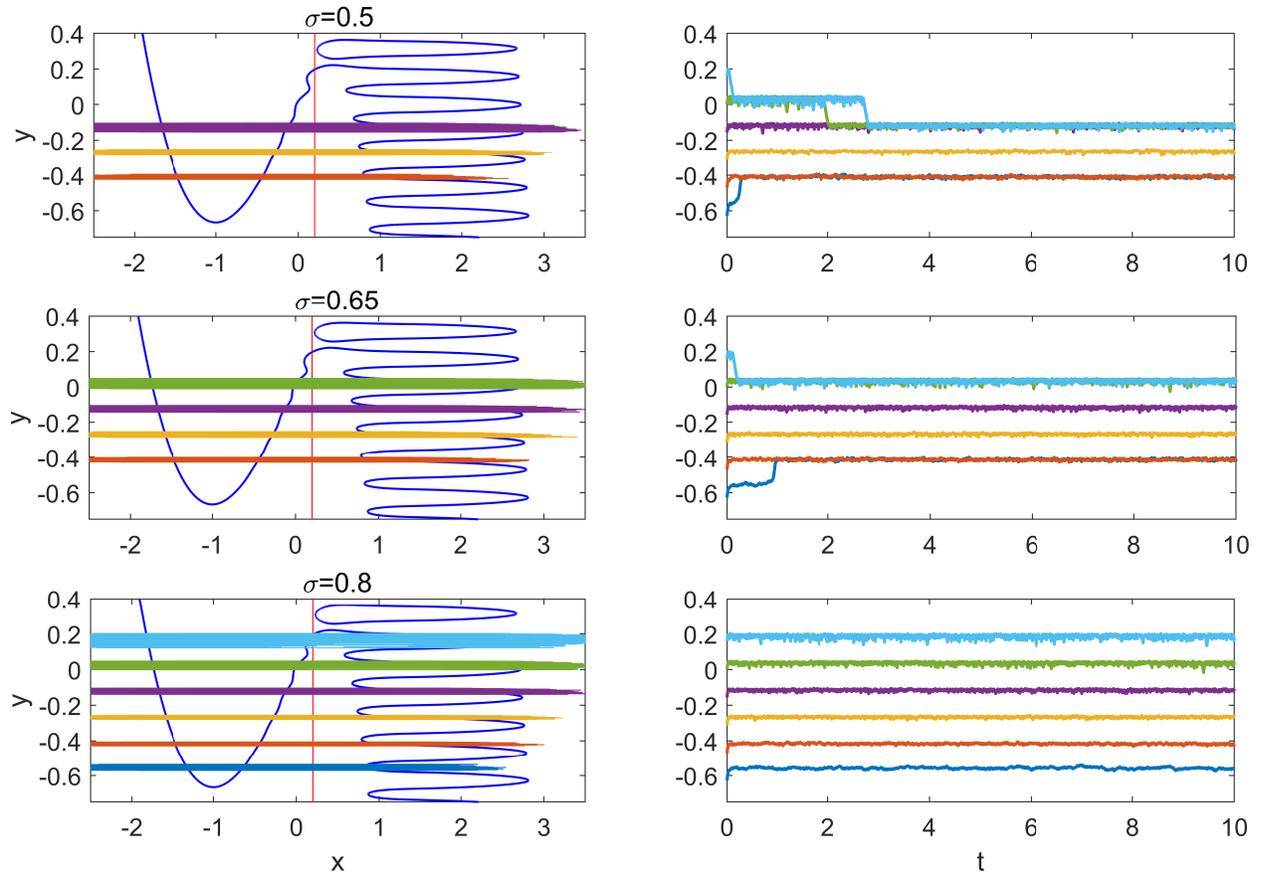}
	\caption{Slow variable traps for large noise. From top to bottom, $\sigma=0.5, 0.65$, and $0.8$. Left panels show the stable phase portraits and orbits; right panels show the time series from five different initial conditions.}
	\label{fig:7}       % Give a unique label
\end{figure}

For the Hedgehog burster investigated in this paper, the non-monotonic potential difference enables more than one stable regions where the slow variable $y$ is restricted in a small interval, which we call slow variable traps. Due to the large noise strength, the transition occurs in a small window of the slow variable $y$ (below the small window, the transition from the left branch is instant and above it the transition from the right branch is instant). In Figure~\ref{fig:7}, the slow variable trap phenomenon is shown for three large values of the noise strength. For $\sigma=0.5, 0.65,$ and $0.8$, there are 3, 4, and 6 stable slow variable traps, respectively.

The widths of the orbits, i.e., the fluctuations of the slow variable in the traps are slightly different from each other. The thinnest orbit in the left panel of Figure~\ref{fig:7} is the yellow one near $y=-0.25$, which corresponds to the first intersection point of the potential difference curves $dU_{ml}$ and $dU_{mr}$ in Figure~\ref{fig:2}B (or the crossing of the transitions positions on different branches in Figure~\ref{fig:4}A). For upper traps, the potential difference $dU_{ml}$ is large and it requires longer interval for the transition from the left branch to occur; while for lower traps, the potential difference $dU_{mr}$ is large and it requires longer interval for the transition from the right branch to take place. At the middle trap at $y\approx-0.25$, the transitions from the left and right branches are balanced, resulting in smaller fluctuations in $y$.

In Figure~\ref{fig:7}, it is interesting to note that before the state becomes trapped into the stable region, the state remains in another metastable region for certain time in the case with $\sigma=0.5$ (cyan and green). This kind of transient trap phenomenon may also occur in coupled SISR systems and influence the collective dynamical behavior in large ensembles.

\section{Discussion}
We have investigated the noise-tuned bursting in a Hedgehog burster, where the noise is applied to the fast variable. Through the proposed distance matching condition, the stochastic periodic orbits were well predicted and the results of Monte Carlo simulations were reproduced. It was found that increasing the noise strength leads to the shrinking of the orbits. This implies that the spike counts in each burst can be tuned by varying the applied noise strength. Moreover, mixed-mode oscillations with the double-escape transition from the right branch can be realized near the boundaries between different stages. Finally, large noise-induced slow variable traps were analyzed, where the number of stable traps depends on the noise strength. The coexistence of multiple traps for the Hedgehog burster is a notable phenomenon that cannot be observed in the FHN model that permits only one trap when large enough noise is applied on the fast variable.

It should be noted that although the Hedgehog burster considered in this paper is in the oscillatory case, noise-tuned bursting can be also investigated in the excitable situation as shown in Figure~\ref{fig:8} (see Appendix \ref{appendix:A}). In that case, coherent stochastic oscillations require a lower bound of the noise strength which depends on the bifurcation parameter $a$. Larger values of $a$ lead to higher transition positions on the left branch as shown in Figure~\ref{fig:8} which illustrates that the bifurcation parameter in the excitable Hedgehog burster will also have a significant impact on the noise-tuned bursting.

Despite that the Hedgehog burster studied in this paper is quite artificial, it shares similar dynamics with other high-dimensional bursting neurons, such as the Hindmarsh-Rose neuron model. It should be noted that the SISR mechanism allows coherent oscillations to occur away from local bifurcations, which leads to robustness of the phenomena investigated in this paper to parameter variations. Therefore, it is promising to observe these phenomena in other fast-slow dynamical systems experimentally. Besides, neurons behave collectively \textit{in vivo}. It has been shown that SISR can play an important role in the interacting excitable FitzHugh-Nagumo systems with synchronization and anticoherence~\cite{Touboul2020}. As synchronization manifests its significance in both normal and pathological cases (for the latter e.g. in Parkinson’s disease and epilepsy), the control of bursting patterns on a single neuron and the ensembles related to the contents investigated in this paper is of great interest. These open problems will be our future works.

\section*{Conflict of Interest Statement}
The authors declare that the research was conducted in the absence of any commercial or financial relationships that could be construed as a potential conflict of interest.

\section*{Author Contributions}
JZ conceived the study, conducted the analysis and simulations, and wrote the manuscript. HN conceived the study and wrote the manuscript. All authors contributed to the article and approved the submitted version.

\section*{Funding}
J.Z. acknowledges support from JSPS KAKENHI JP20F40017, Natural Science Foundation of Jiangsu Province of China (BK20190435) and the Fundamental Research Funds for the Central Universities (No.30920021112). H.N. thanks JSPS KAKENHI JP17H03279, JP22K11919, JP22H00516, JPJSBP120202201, and JST CREST JP-MJCR1913 for financial support.

\section*{Data Availability Statement}
The raw data supporting the conclusions of this article will be made available by the authors, without undue reservation.

\appendix
	
	\section{Noise-induced and noise-tuned bursting in excitable Hedgehog burster}
	\label{appendix:A}
	
		Although the Hedgehog burster investigated in this paper has a stable limit cycle without noise, we have shown that the stochastic periodic orbits induced by SISR are different from the deterministic limit cycle. In fact, even when there is no deterministic limit cycle in the Hedgehog burster, i.e., when the bifurcation parameter $a>1$ in system (\ref{eq:1}), noise can still induce coherent oscillations.
	
\begin{figure}
	\centering
	\includegraphics[width=1\textwidth]{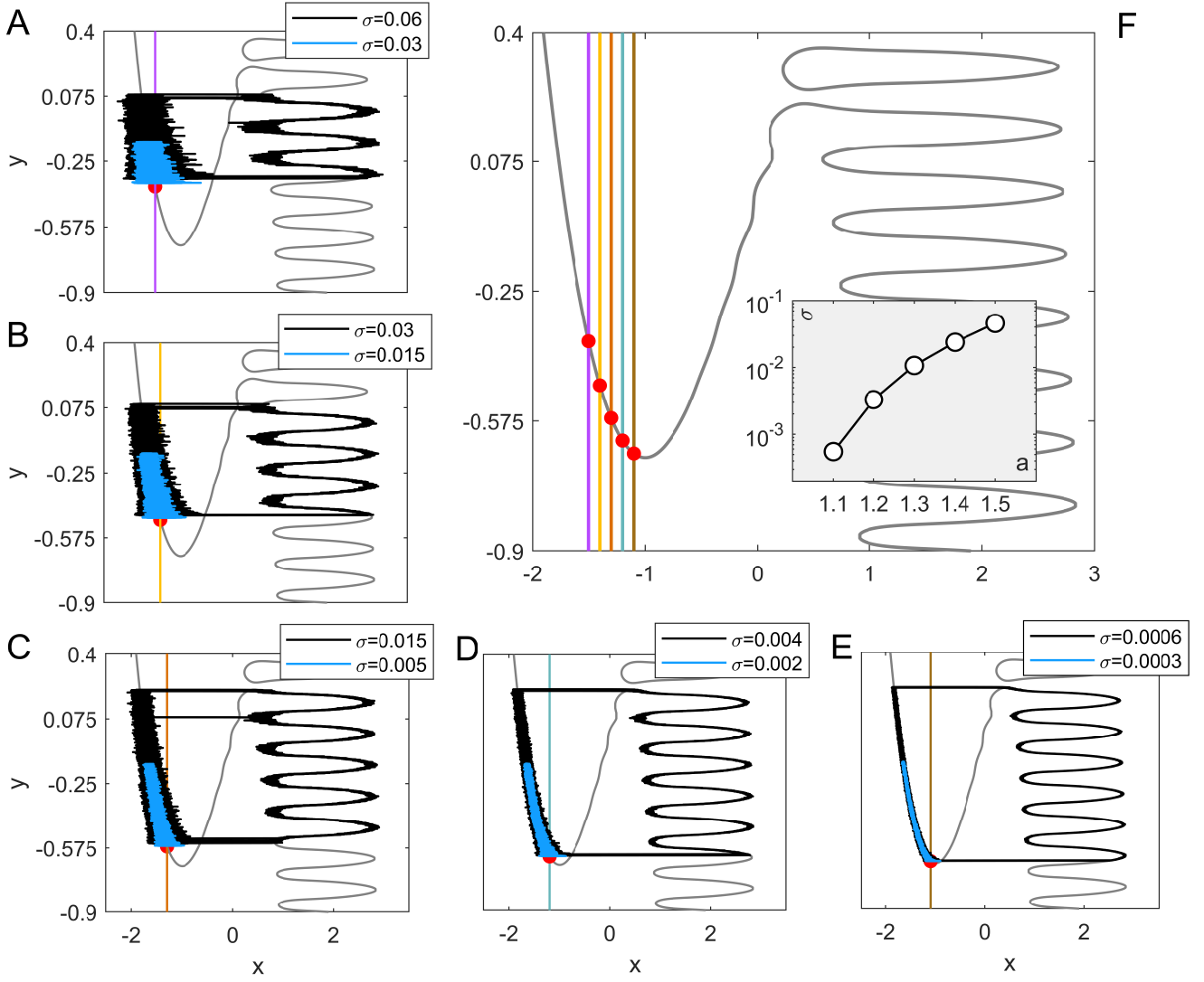}
	\caption{Noise-induced and noise-tuned bursting in excitable Hedgehog bursters. (A)-(E) The bifurcation parameters are: $a=1.5, 1.4, 1.3, 1.2$, and $1.1$, respectively. Black and blue curves represent trajectories for noise strength above and below the threshold values shown in the inset of (F). Red dots denote the stable equilibrium points. (F) Variation of the $y$ nullcline for different bifurcation parameter $a$. The inset displays the threshold values of noise strength obtained from the distance matching condition (\ref{eq:5}).}
	\label{fig:8}       % Give a unique label
\end{figure}

		Different from the oscillatory case, the excitable Hedgehog burster has a lower bound of the noise strength \cite{Zhu2021PRR,Yamakou2018}, below which the transition to the right branch will be a Poisson process. The lower bounds of noise strengths for different values of $a$ can be calculated from the distance matching condition (\ref{eq:5}) when the lhs and rhs have no intersection point. The critical values are plotted in the inset of Figure~\ref{fig:8}(F). As expected, the lower bound increases with $a$ since an earlier transition requires a larger noise strength. Figures~\ref{fig:8}(A)-(E) illustrate stochastic trajectories above and below the lower bound of noise strengths, where the former can induce coherent oscillations while for the latter noise is too weak to initiate transitions.
	
		The noise-tuned bursting behaviors can be similarly investigated as in the main text. The difference is that there is a lower bound of noise strength in the excitable case in contrast to the oscillatory case. Coherent oscillations can only be observed above that bound. Therefore, the spike counts in each burst can be also influenced. For example, there are at most five spikes for $a=1.3$ (Figure~\ref{fig:8}(C)) whatever the noise strength is. As a result, in the excitable Hedgehog burster, the bifurcation parameter $a$ can also have significant influences on the noise-tuned bursting. On the one hand, the transition position on the left branch can be raised by the equilibrium point. On the other hand, the lower bound of noise strength for coherent oscillations will also modify the transitions on the right branch.

\clearpage


%%% Make sure to upload the bib file along with the tex file and PDF
%%% Please see the test.bib file for some examples of references

%%% If you are submitting a figure with subfigures please combine these into one image file with part labels integrated.
%%% If you don't add the figures in the LaTeX files, please upload them when submitting the article.
%%% Frontiers will add the figures at the end of the provisional pdf automatically
%%% The use of LaTeX coding to draw Diagrams/Figures/Structures should be avoided. They should be external callouts including graphics.

\end{document}